\documentclass[twocolumn,amsmath,amssymb,amsfonts]{revtex4-1}

\usepackage{graphicx}
\usepackage{dcolumn}
\usepackage{bm}
\usepackage{latexsym,color}

\usepackage{braket}

\usepackage{hyperref}

\hypersetup{
  colorlinks,
  citecolor=blue,
  linkcolor=blue,
  urlcolor=blue}

\newcommand{\be}{\begin{equation}}
\newcommand{\ee}{\end{equation}}

\newcommand{\ba}{\begin{eqnarray}}
\newcommand{\ea}{\end{eqnarray}}

\begin{document}

\title{Nonlinear optical conductance from photon-assisted tunneling\\ in a nanoantenna-coupled tunnel diode}
\author{Paul S. Davids$^{1}$}\email{pdavids@sandia.gov}\
\author{Erik Anderson$^{2}$}
\author{Jared Kirsch$^{1,3}$}
\author{Joshua Shank$^1$}

\affiliation{%
$^{1}$Sandia National Laboratories, Albuquerque, NM 87185 \\
$^2$Department of Mechanical Engineering, Georgia Tech,   Atlanta, GA 30332\\
$^3$Department of Mechanical Engineering, University of New Mexico,  Albuquerque, NM 87131
}%
\date{\today}

\begin{abstract}
Ultrafast photon-assisted tunneling is shown to lead to large nonlinear dynamic currents that can be used to generate dc rectified current as well as higher order harmonic response from a metasurface-coupled tunnel diode.   This artificial nonlinear media can have tailored resonances in the infrared around the longitudinal optical phonon modes of various compatible restrahlen oxides leading to enhanced electric field confinement in a metal-oxide-semiconductor tunnel diode.   This provides a new avenue for the examination and design  of nonlinear phenomena extending the concept of  nonlinear susceptibility to nonlinear conductance and the dynamic current in these devices.
\end{abstract}


\maketitle


Maxwell's equations form the basis of all light matter interactions,  
\begin{equation}
\begin{array} {clc}
\nabla \times \mathbf{ E}  = -\partial_t \mathbf{ B} & ,& \nabla \cdot \mathbf{ D} = \rho \\
\nabla \times \mathbf{ H} =  \mathbf{ J} + \partial_t \mathbf{ D}&,& \nabla \cdot \mathbf{ B} = 0, 
\end{array}
\label{disp:eq1}
\end{equation}
where the derived fields, $ \mathbf{D} = \epsilon_0 \mathbf{E} + \mathbf{P}$ and $\mathbf{B }= \mu_0(\mathbf{H}+\mathbf{M})$, relate  the fundamental electromagnetic fields, $(\mathbf{E,H})$, to the macroscopic polarization and magnetization response of the media\cite{panofsky2012classical}. Constituent relations describe light-matter interaction in the classical field equations and incorporate quantum effects into the underlying material response to the electromagnetic fields. These constituent relationships are inherently connected  to  microscopic models which relate the averaged local current, charge polarization, and magnetization distributions to the electric and magnetic fields.   Fundamentally, Maxwell's displacement current arises from the requirement of charge conservation, $\nabla \cdot \mathbf{J} + \partial_t \rho  = 0$. The total current,  $\mathbf{J}_{tot} = \mathbf{J} + \partial_t \mathbf{D}$, consists of the conduction current $\mathbf{J}$  and  displacement current  $ \partial_t \mathbf{D}$,  which generates  the linear and non-linear coupling of electric and magnetic fields to matter.
Thus a time varying polarization charge distribution in a dielectric or metallic structure gives rise to a displacement current that can lead to engineered resonances and tailorable optical response.  
The total current in the frequency domain, $\mathbf{J}_{tot}(\omega) = \mathbf{J}(\omega) -i \omega\epsilon_0 \mathbf{E}(\omega) -i\omega \mathbf{P}(\omega)$, therefore depends on the dynamic polarization and conduction current. In standard non-linear optics\cite{Boyd:2008:NOT:1817101},   the polarization is typically expanded in a  power series in the electric field to give
\be
P_i = \epsilon_0 \left( \chi^{(1)}_{i,j} E_j +  \chi^{(2)}_{ijk} E_j E_k + \chi^{(3)} _{ijkl} E_j E_k E_l \ldots \right) ,
\ee
where $\chi^{(i)}$ is the non-linear susceptibility, and the Einstein summation convention is used. The frequency-dependent dielectric constant is given as $\epsilon(\omega) = (1+ \chi^{(1)}(\omega))$ for isotropic media and forms the basis for the linear optical response of most materials. The higher order non-linear susceptibilities generate many interesting phenomena such as second and third harmonic generation, sum and difference frequency generation, optical phase-conjugation and many more.  Furthermore, these non-linear interactions can be used in parametric processes for amplification and non-linear spectroscopies and play a central role in ultrafast quantum optics. 

\begin{figure*}[!htbp]
\includegraphics[width=0.9\textwidth]{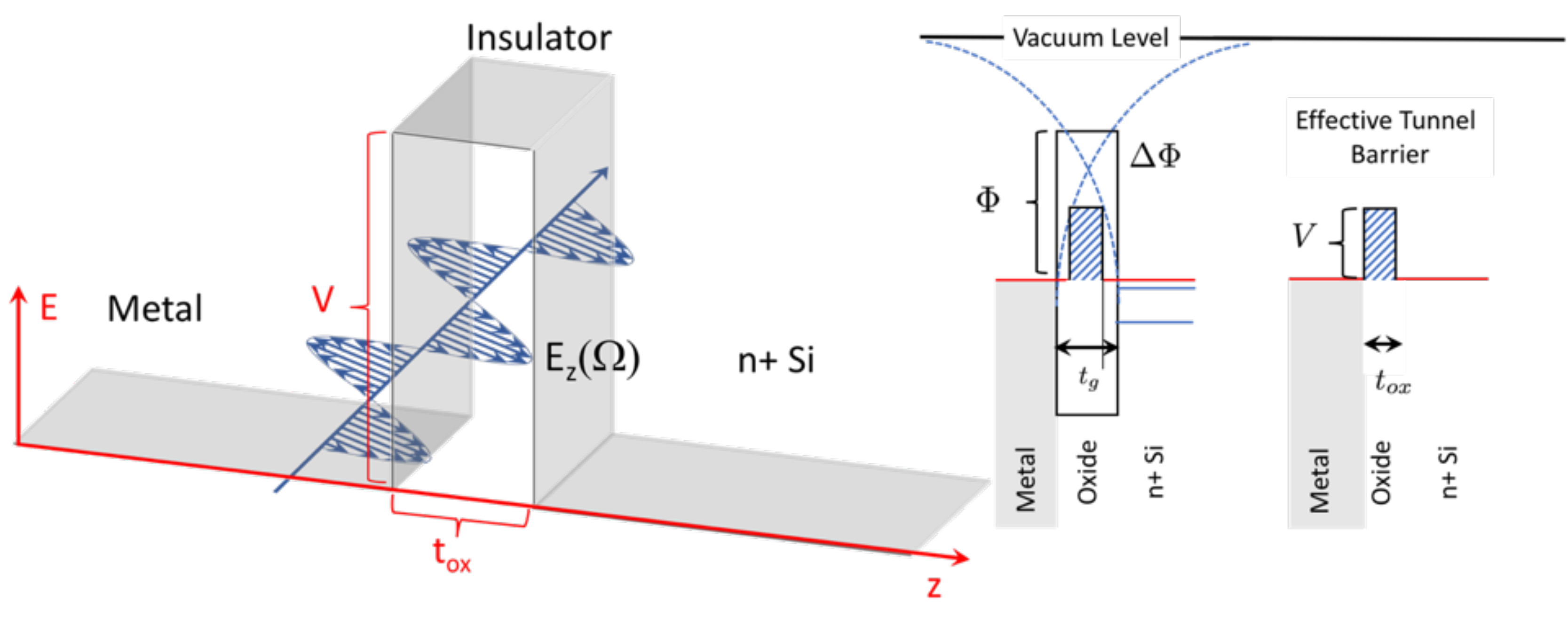} 
\caption{Schematic view of photon-assisted tunneling in a metal-oxide-n+ silicon tunnel diode. The band diagram shows the effects of image force lowering and illustrates the reduction in barrier height and physical thickness in the uniform barrier model.}
\label{fig:1}
\end{figure*}

Recently, we have developed a model for ultra-fast dynamic tunneling based on a density matrix expansion of the transfer Hamlitonian\cite{PhysRevB.97.075411}.  The transfer Hamiltonian approach is formulated as two half-spaces separated by a thin barrier interaction that connects the two half-spaces such that we can describe the system phenomenologically as $H = H_l + H_r + H_T(t)$, where r and l refer to right and left half-spaces\cite{Bardeen_1961,harrison,Prange}.  The photon-assisted tunneling transfer interaction is  the electromagnetic dipole interaction defined in the tunnel barrier and is
\be
H_T(t) =  \int_{-\infty}^{\infty} \frac{d\omega}{2\pi} e^{-i\omega t} \frac{e\hbar}{m \omega} \mathbf{E}(\omega) \cdot \nabla,
\label{eq:transfer}
\ee
where we have used the Fourier transform of the vector potential  to derive the electric field along with the constraints of real valued fields.  The quantum mechanical momentum in the barrier is purely imaginary and the wavefunctions exponentially decays into the barrier from the left and right half spaces.  The transfer Hamiltonian matrix elements formed from the overlap of the exponentially decaying wavefunctions  determines the tunneling probability drives the photon-assisted tunneling.  The transfer Hamiltonian for the confined electromagnetic interaction is anti-Hermitian, $H^*_T(t) = - H_T(t)$,

The tunneling current is expanded  in terms of the density matrix is where $\rho_{nm}$ is the density matrix and $H^T_{nm}$ is the complex transfer Hamiltonian matrix elements evaluated in the tunnel barrier region.  The density matrix  power series in the transfer interaction strength, $\lambda$, is 
\be
\rho_{nm} = \rho^{(0)}_{nm} +  \lambda \rho^{(1)}_{nm} +  \lambda^2 \rho^{(2)}_{nm}  +\ldots,
\ee
where $\lambda=1$ represents the fully interacting system.  In this model we have shown that the density matrix expansion leads to an effective non-linear expansion of the conduction current in the device. Explicitly,  the density matrix expansion in the transfer Hamiltonian model for tunneling in these structures leads to an expansion of current density
\be
J_i = \sigma^{(1)}_{ij} E_j + \sigma^{(2)}_{ijk} E_j E_k +\sigma^{(3)}_{ijkl} E_j E_k E_l+ \ldots
\label{eq:nlcurrent}
\ee
where the non-linear nature of the tunneling current as a function of transverse field in the gap generates the non-linear conductances in a similar fashion  to the dynamic polarization of a media leads to  the non-linear susceptibility.   The non-linear conductances are completely analogous to the non-linear susceptibilities from standard non-linear optics and arise  from multi-photon tunneling processes at each order in the expansion\cite{PhysRevB.97.075411,PhysRevLett.73.3443,PhysRevLett.75.4098,Tien_Gordon_multiphoton}. The only difference is that these conductances arise from ultra-fast direct tunneling in an artificially structured photonic media. 

The paper is organized as follows. In section \ref{sec1}, the generalized dynamic tunneling current is examined through third order in the density matrix expansion.  The nonlinear form-factors are defined in terms of matrix elements of the transfer interaction which is proportional to the confined electric field in the gap.  Section \ref{sec2} examines the nonlinear expansion in the uniform barrier model.  In this model, the explicit matrix elements are evaluated and a parametric study of the barrier height and thickness are examined for a frequency dependent transverse electric field.   The  second order dynamic current in the uniform barrier model is the main focus of this section  and is examined parametrically.  The special case of thermal illumination is reviewed and the current versus temperature as a function of barrier height and barrier thickness is examined.   Moreover, the effect of an applied bias on the metal is approximated within the uniform barrier picture and current voltage characteristics are obtained.   In section \ref{sec3}, the case of a monochromatic transverse optical field is examined to third order in the expansion.  The first order current gives the tunneling optical conductivity at the driving frequency of the  optical transverse field.  The second order current expression gives rise to a dc conductivity and a large second harmonic response as expected from the measured response of the grating-coupled tunnel diode rectifier\cite{Davids:2015aa,kadlec_prapplied,PhysRevApplied.9.054040}.  Finally, the third order current expression gives rise to a correction to the first order linear response and a third harmonic term.  
In section \ref{sec4}, we present a summary of the findings and examine the prospect of measuring higher order harmonics in grating-coupled tunnel diode rectifiers and possible new devices for mixing and generation of THz signals  in metallic coplanar waveguides.

\section{Dynamic Tunneling}\label{sec1}

The dynamic current  density can be obtained from the density matrix expansion and  in the transfer Hamiltonian formalism described previously\cite{PhysRevB.97.075411}.  The frequency-dependent current density can be written as a power series, $J(\omega) = J^{(1)}(\omega) + J^{(2)}(\omega)+J^{(3)}(\omega)\ldots$,  corresponding to the strength of the transfer interaction for a generally dispersive transverse electric field and is given by
\begin{widetext}
\ba 
J^{(1)}(\omega)  & =  & i\frac{2e}{\hbar} \left(\frac{e\hbar  E_z(\omega)}{m \omega}\right)F^{(1)}\ \label{j1}\\
J^{(2)}(\omega)    & = &i\frac{2e }{\hbar} \int\limits_{-\infty}^{\infty} \frac{d\Omega}{2\pi}  \left(\frac{e\hbar E_z(\Omega)}{m \Omega}\right) \left(\frac{e\hbar  E_z(\omega-\Omega)}{m(\omega- \Omega)}\right) 
  F^{(2)}(\omega-\Omega)  \label{j2}\\ 
 J^{(3)}(\omega)  & = & i\frac{2e}{\hbar} \int\limits_{-\infty}^{\infty} \frac{d\Omega}{2\pi}  \int\limits_{-\infty}^{\infty} \frac{d\omega'}{2\pi} \left( \frac{e\hbar E_z(\Omega)}{m \Omega} \right)  \left( \frac{e\hbar  E_z(\omega') }{m \omega'}\right)  \left(\frac{e\hbar E_z(\omega-\Omega-\omega') }{m (\omega-\Omega-\omega')} \right)   F^{(3)}(\omega-\omega',  \Omega)  \label{j3} \\
 \vdots & &  \vdots \nonumber
\ea
\end{widetext}
where $F^{(i)}$  are the frequency-dependent form factors. Here $E_z(\omega)$ is the frequency-dependent transverse electric field in the tunnel gap.  The transverse electric fields are real-valued such that $E_z(-\omega) = E^*_z(\omega)$. These form-factors relate the matrix elements of the momentum in the tunnel barrier to  energy denominators arising from the density matrix expansion factors are 
\ba
F^{(1)} & = & \frac{1}{A}  \sum_{nm} u_{nm} f_n \delta_{n,m}  \nonumber \\
F^{(2)}(\omega_1)  & = & \frac{1}{A}    \sum_{nm} u_{mn} u_{nm} \big(f_m - f_n \big)G_{nm}(\omega_1)  \nonumber \\ 
F^{(3)}(\omega_1,\omega_2)  & = &  \frac{1}{A} \sum_{lnm} u_{mn}u_{nl}u_{lm} G_{nm}(\omega_1) \left[ (f_{m} - f_{l}) \times  \right. \nonumber \\
& &  \left.   G_{lm}(\omega_2)  -G_{nl}(\omega_2)  (f_{l} - f_{n}) \right]   \label{ff1} ,        
\ea
where $A$ is the tunnel diode surface area, $u_{nm} = \braket{n|\nabla_z m}$,  $\rho^{(0)}_{nm} = f_n\delta_{nm}$ is the zeroth-order static quasiparticle distribution function, and $G_{nm}(\omega) =(\hbar \omega -E_n +E_m)^{-1}$ is the non-interacting propagator.  These non-linear form factors and the associated dynamic current densities can be evaluated explicitly  in the context of a particular model. This will be presented for the uniform barrier model outlined in the following section and appendix \ref{app:1}.

\begin{figure*}
\includegraphics[width=1.0\textwidth]{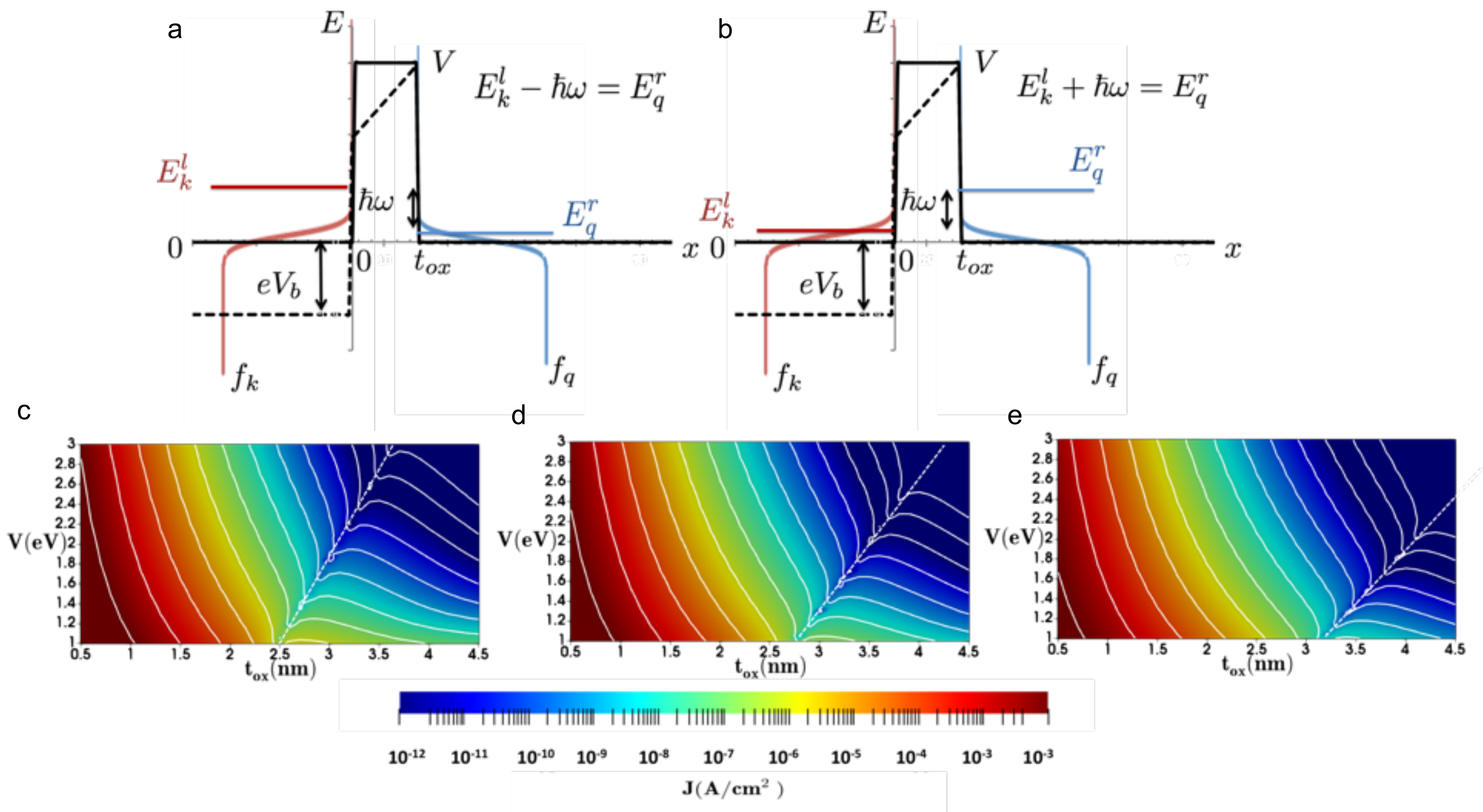} 
\caption{a. Schematic of photon-assisted tunneling from semiconductor to metal. b. Photon-assisted tunneling from metal to semiconductor.  Contour plots of zero bias photon-assisted DC tunnel current as a function of barrier height $V$ and width $t_{ox}$ for differing source temperatures. c. Source temperature  T=400 $^0$C d. Source temperature  T=300 $^0$C e. Source temperature  T=200 $^0$C.  The dashed contours indicate the approximate locus of points where current density changes sign.}
\label{fig:2}
\end{figure*}

 \section{Photon-assisted Tunnneling: Uniform Barrier}\label{sec2}
The uniform barrier photon-assisted tunneling problem is illustrated in figure \ref{fig:1}  and neatly corresponds to aluminum-SiO$_2$-n+ silicon (nMOS) tunnel diode.  In this device,  the aluminum work-function  and the conduction band of silicon are aligned and separated by a uniform thin (3-5 nm) thick tunnel oxide barrier.  This band alignment implies that there is not a built in field in the oxide  resulting in a uniform square tunnel barrier. Image force lowering of the barrier can  reduce the barrier height and shorten the tunneling length from the physical oxide thickness to an effective oxide thick, $t_{ox}$. 
Figure \ref{fig:1} shows that the optical field is confined in the insulating tunnel barrier between the two electrodes.   This type of field confinement has been shown to arise from coupled photon-phonon resonances resulting in large enhancement of optical fields in extreme subwavelength gaps\cite{Alam795,alu2007epsilon,kadlec_prapplied,PhysRevApplied.4.044011,PhysRevApplied.9.054040}.  This enhanced confinement gives rise to large photon-assisted tunneling and rectification due to the asymmetry between the semiconductor and metal density of states\cite{chen2016giant}.  

The dynamic current density expansion through 3rd order  is given by Eq. \ref{j1}-\ref{j3}  where Eq. \ref{ff1} gives  explicitly the general expression  frequency dependent form-factors.  Previously, we have shown that the 2nd order expansion for the tunneling current is a generalized form of Bardeen's transfer Hamiltonian tunneling current.   In the following, we examine the 2nd order dynamic tunneling current and look at the real-valued dc current density.  This current density is associated with direct rectification of the optical transverse field through photon-assisted tunneling.   
The 2nd order dynamic current density is in general complex in the presence of a dispersive frequency dependent field in the gap.  By considering the imaginary part of the form-factor, we can obtain the 2nd order expression that leads to a real-valued dc rectified tunnel current.   The resulting current density can be evaluated for the uniform barrier and   is 
\begin{widetext}
\ba
J^{(2)} (\omega) & = & 2\pi AT^2 \left( \frac{m_k m_q}{m^2} \right)  \int\limits_{-\infty}^{\infty} \frac{d\Omega}{2\pi}  \left( \frac{eE_z(\Omega) t_{ox} }{\hbar \Omega} \right) \left( \frac{eE_z(\omega-\Omega)t_{ox}} {\hbar (\omega- \Omega)} \right)\beta \int\frac{dE_k}{2\pi} \tau^2(E_k,E_k + \hbar\Omega -\hbar \omega) \times \nonumber \\
&  &  \ln\left[ \frac{ 1 + \exp(-\beta( E_k+\hbar\Omega -\hbar \omega))}{ 1 + \exp(-\beta E_k) } \right]  .
\ea
\end{widetext}
Here $A$ is Richardson's coefficient, $T$ is the  source temperature, $\beta = 1/k_bT$ and $\tau(E_l,E_r)$ is the left-right tunneling transmission through the barrier. . There is also an imaginary contribution from the frequency dependent form-factors that comes from the principal integral part of the Green's function.
The incoherent rectified dc current density for the case of dispersive transverse fields in the tunnel gap is obtain by considering  $\omega=0$, we have 
\begin{widetext}
\ba
J^{(2)}_{dc} & = & -AT^2 \left( \frac{m_k m_q}{m^2} \right)  \int\limits_{-\infty}^{\infty} \frac{d\Omega}{2\pi}  \left( \frac{e|E_z(\Omega)| t_{ox}}{\hbar \Omega} \right)^2 \beta \int dE_k~ \tau^2(E_k,E_k + \hbar\Omega ) \ln\left[ \frac{ 1 + \exp(-\beta( E_k+\hbar\Omega )}{ 1 + \exp(-\beta E_k) } \right] , \label{eq:jdc} 
\ea
\end{widetext}
where $A$ is Richardson's constant, $T$ is the carrier temperature, $\beta=1/k_b T$, and $\tau(E,E+\hbar \Omega)$ is the tunnel barrier transmission.  If we consider blackbody illumination of the device, we can approximate the field amplitude in the tunnel barrier.   
The power per unit area per unit bandwidth emitted from a blackbody is,
$M_0(\nu,T) \simeq |E_0(\Omega)|^2/2Z_0$,
where $Z_0 = \sqrt{\mu_0/\epsilon_0}$ is the impedance of free-space, $\Omega = 2\pi\nu$, and $M_0(\nu,T)$ is the spectral exitance of a blackbody in free-space. 

The incoherent field amplitude in the tunnel barrier is enhanced relative to the incident field due to the nanoantenna resonance and the ENZ response of the oxide barrier material.  The  enhanced field in the tunnel barrier is  $E_z(\Omega) = \gamma E_0(\Omega)$, and  is
\be
|E_z(\Omega)|^2 =  2 Z_0 \gamma^2\left(\frac{\hbar \Omega^3}{2\pi c^2}\right) \frac{ 1}{\exp(\beta \hbar \Omega) - 1}.
\ee
where $\gamma = E_z/E_0$ is the field enhancement factor which can be estimated from EM simulations.


The numerical evaluation of the analytic dc current density in the uniform barrier model given by Eq. \ref{eq:jdc} requires the the tunnel barrier height and width. For all simulations,   the effective longitudinal electron masses for (100) silicon is $m_{si} = 0.19 m_e$ and the metal is $m_m = m_e$\cite{sze2006physics}.  The band diagram relative to the vacuum level uses the work function of aluminum as 4.2 eV (4.06-4.26 eV) and the electron affinity of silicon as approximately 4.05 eV\cite{sze2006physics}. The bulk oxide barrier height relative to the metal is 3.0 eV with a bandgap of 8.0 eV.   The integral over the frequency of the  transverse field in the tunnel barrier can be approximated by the integrand evaluated at the resonant frequency corresponding to a wavelength of $\lambda$ = 7.5 $\mu$m, and multiplied by the bandwidth of the resonance, $\Delta \nu$ = 1 THz.  The field concentration factor is arbitrarily held fixed at $\gamma = 100$.

\begin{figure*}
\includegraphics[width=0.9\textwidth]{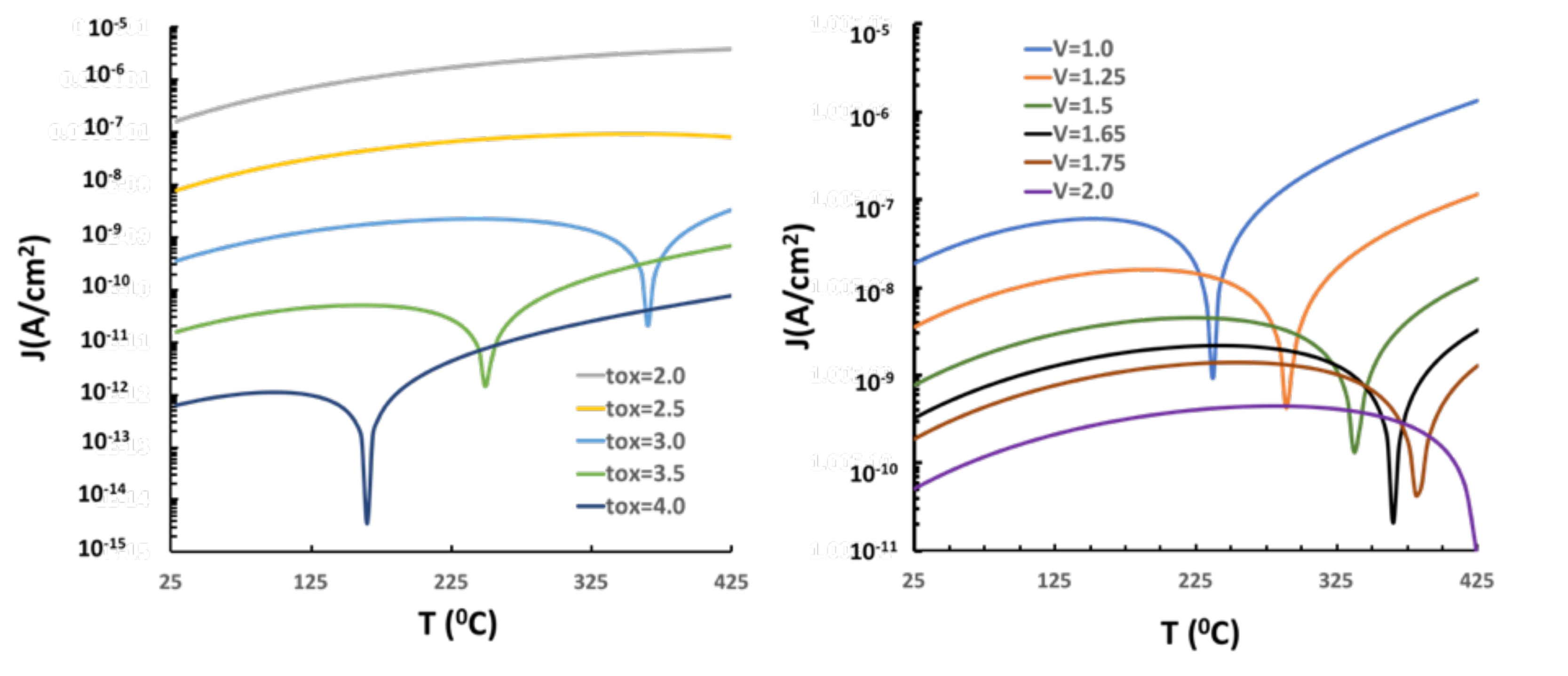} 
\caption{ Photon-assisted current density in the uniform barrier model as a function of the source temperature of the transverse electric field in the gap.  (a) The current density for fixed barrier height $V=1.65$ eV  for different uniform tunnel barrier oxide thicknesses. (b) The current density for a fixed uniform tunnel barrier thickness, $t_{ox} = 3$ nm, and variable barrier height $V$.    }
\label{fig:3}
\end{figure*}

Furthermore, the integral can be restricted to positive frequencies which results in two terms; a  photon emission term and a photon absorption term.  Figure \ref{fig:2}(a) and (b) illustrates  photon emission and photon absorption processes, respectively.  In the emission process, an electron in an occupied state in the semiconductor near the Fermi energy absorbs a photon and tunnels to an unoccupied state in the metal.  The time-reversed process shown in Fig.\ref{fig:2}(b) is the result from absorption of the photon in the metal and tunneling into an unoccupied state in the semiconductor.  These terms contribute with opposite signs since the 2nd order current expression is proportional to the difference in the occupation of the left and right sides.  The difference in the density of states between the semiconductor and metal results in the tunneling asymmetry.  This can be simply seen as arising from the difference in the transverse effective masses in the Si and the metal. (See definition of $F^{(2)}$ in Eq. \ref{ff1})  Figure \ref{fig:2} (c)-(e) show parametric studies of the the photon-assisted tunnel current as a function of barrier height and barrier thickness for various source temperatures.   A locus of sign change in the current occurs which is a direct result of the cross over between photon emission and photon absorption previously mentioned.   At small barrier height and small barrier thickness, the tunneling current can be quite large and it should be noted that we have ignored the contribution of thermionic emission over the barrier.   At larger barrier height and thickness, we see a drop off in the current in general, and a change in sign depending on which side of the dashed line the parameters fall on.  Figure \ref{fig:3} (a)  and (b) shows the parametric plots of the 2nd order uniform barrier tunnel current as the  source temperature  increased for fixed barrier height and oxide thickness, respectively.  In these plots we see that the  change in sign occurs at a characteristic temperature determined by barrier height and thickness.  

\begin{figure*}
\includegraphics[width=0.8\textwidth]{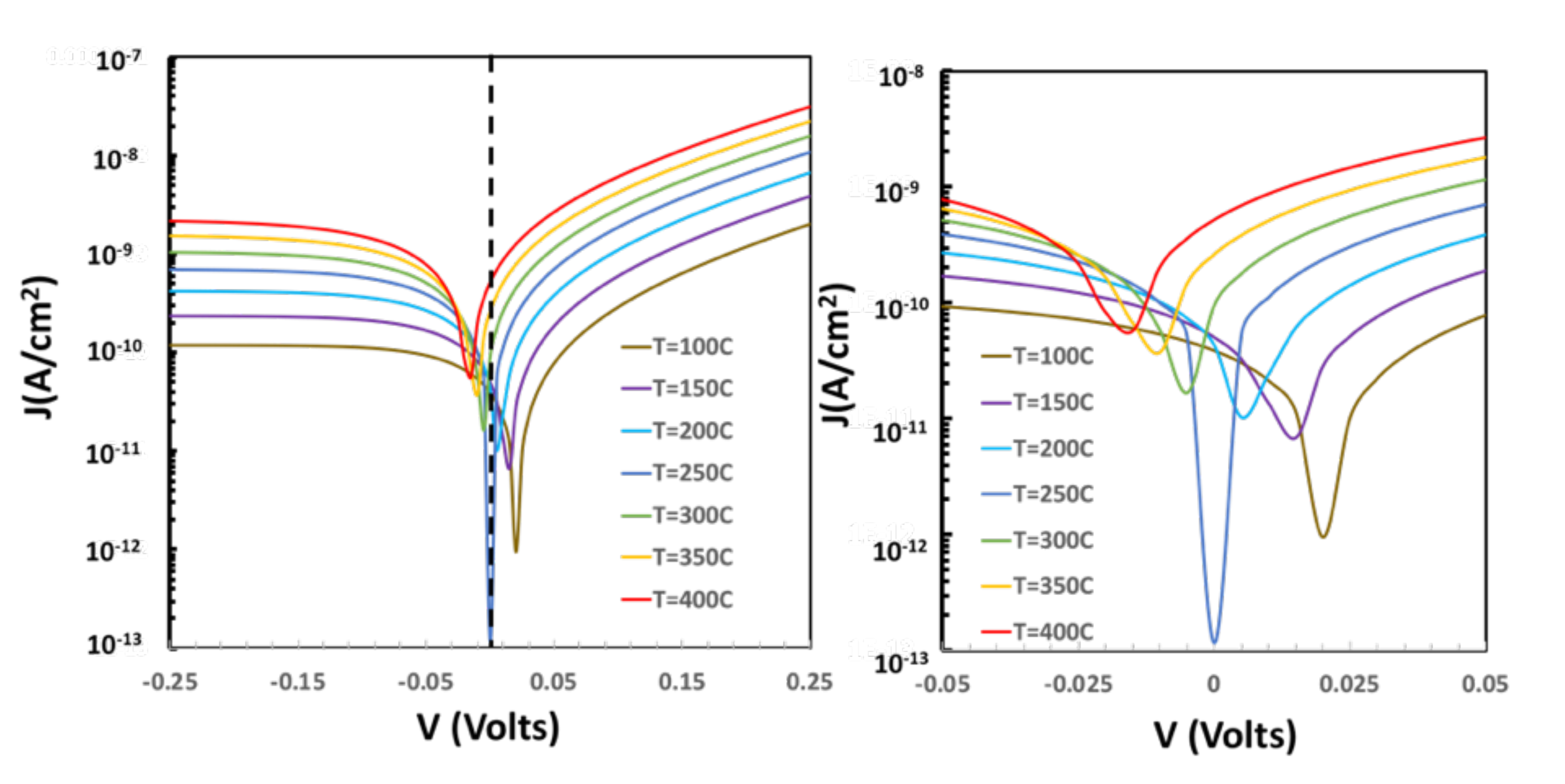} 
\caption{Current voltage characteristics  for various source temperatures. The barrier height is $V=1.65$ eV and the barrier width $t_{ox}=3.5$ nm. Right plot is IV characteristic on an reduced scale near $V=0$. }
\label{fig:4}
\end{figure*}

The effect of an externally applied bias can be examined in the uniform barrier model  by changing the effective chemical potential of the metal as shown in Fig. \ref{fig:2}.  This is approximate since we explicitly assume the barrier shape and resulting transmission is unchanged.  The resulting current voltage characteristics can be obtained for the device under bias for various temperatures.   In figure \ref{fig:4},  the minimum in the current on the log plot is clearly visible.  The current minimum separates negative (left of min) from positive (right of min) current flow and is seen to shift from positive to negative voltage as we increase the source temperature of the transverse electric field in the gap. At approximately $T=250^0$C, we find that the current minimum is at zero bias which represents a current balance condition between photon-emission and photon-asbsorption in the photon-assisted tunneling current.  We also note that the asymmetry of the diode characteristics result mainly from the effective longitudinal mass difference.  

\section{ Nonlinear conductances}\label{sec3}

The expansion for the dynamic current density has been derived in the previous section for an arbitrary dispersive transverse field in the tunnel barrier.  For coherent photon-assisted tunneling, we consider  a single frequency field, $E_z(t) = E_0\sin(\omega_0t)$ confined in the tunnel gap where the frequency dependent field is 
$E_z(\omega) = i\pi E_0\left( \delta(\omega-\omega_0) - \delta(\omega+\omega_0)\right)$
where $E_0$ is the real valued amplitude.   The resulting frequency dependent current is 
\begin{widetext}
\ba
J^{(1)}(\omega) & = &   -\frac{2e}{\hbar}  \pi \left(\frac{e\hbar}{m \omega_0}\right)  E_0\big\{ \delta(\omega-\omega_0) +\delta(\omega+\omega_0)\big\}  F^{(1)}  \label{eq:current0} \\
J^{(2)}(\omega)  & = &   -i\frac{2e }{\hbar} \frac{\pi}{2}  \left(\frac{e\hbar}{m \omega_0}\right)^2  E_0^2 \left\{ \big[ \delta(\omega-2\omega_0) F^{(2)}(\omega_0) + \delta(\omega+2\omega_0) F^{(2)}(-\omega_0) \big]  \right. \nonumber \\
& & + \left. \big[F^{(2)}(\omega_0)+ F^{(2)}(-\omega_0)\big]\delta(\omega)  \right\} \label{eq:current1} \\ 
J^{(3)}(\omega)  & = &    +\frac{2e }{\hbar} \frac{\pi}{4}  \left(\frac{e\hbar}{m \omega_0}\right)^3  E_0^3 \left\{   F^{(3)}(2\omega_0,\omega_0) \delta(\omega-3\omega_0)  +F^{(3)}(-2\omega_0,-\omega_0) \delta(\omega+3\omega_0)  \right.  \nonumber  \\
&  & +\delta(\omega -\omega_0)\big[ F^{(3)}(0,\omega_0) +F^{(3)}(0,-\omega_0) +F^{(3)}(2\omega_0,\omega_0) \big] \nonumber      \\      
&   &  +\left.  \delta(\omega +\omega_0)\big[ F^{(3)}(0,-\omega_0) +F^{(3)}(0,\omega_0) +F^{(3)}(-2\omega_0,-\omega_0) \big]  \right\}. \label{eq:current2}
\ea
\end{widetext}
Eq. \ref{eq:current0} - \ref{eq:current2} gives the dynamic current density in powers of the transverse electric field amplitude through third order in the density matrix expansion.   The effective non-linear conductance is obtained by dividing the nth order current density at the  by  the power of the transverse field, $\sigma^{(n)}(\omega)=J^{(n)}(\omega)/E_0^n$. 

The first order current represents the  electron tunneling displacement current through the tunnel junction at the frequency dependence of the transverse field.  The second order current has two contributions; a second harmonic contribution at $\omega=\pm 2 \omega_0$ and a dc term at $\omega = 0$.   The third order term contains the third harmonic at $\omega = \pm 3 \omega_0$, and a linear term at $\omega=\pm\omega_0$ which acts as a small correction to the first order term.   It is important to note that the first and third order currents do not allow for a DC current, only the second order term.

\begin{figure*}[!htb]
\includegraphics[width=0.9\textwidth]{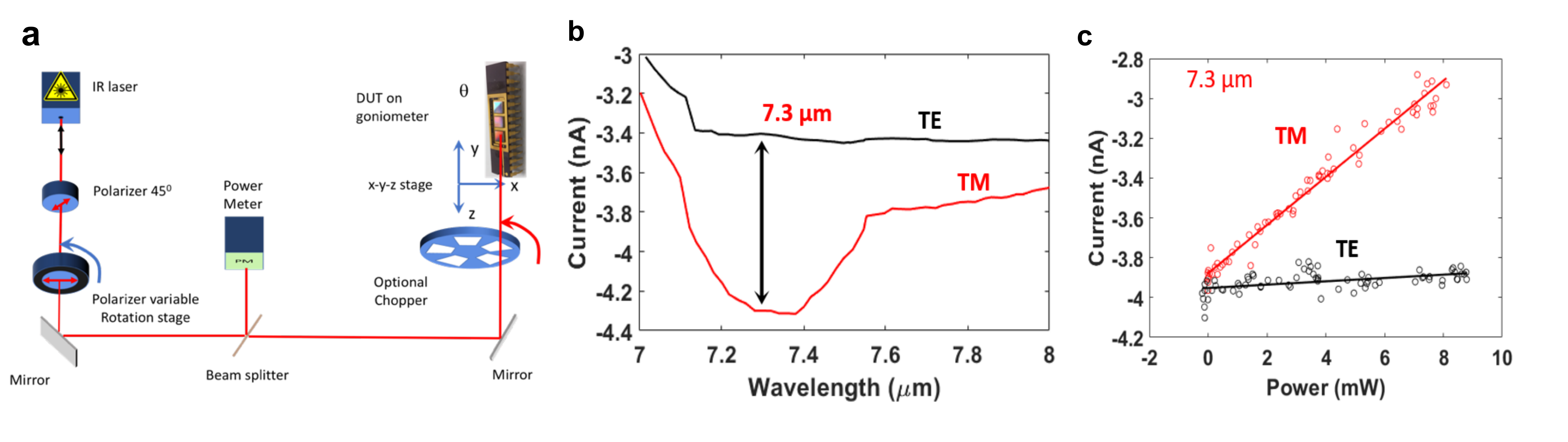}
 \vspace{-0.5cm}
\caption{ Experimental measurement of dc response of grating-coupled unipolar tunnel diode. a) Experimental setup for measurement of photoresponse of infrared device using a tunable quantum cascade laser.  b) Spectral photocurrent response of dc 2nd order photon-assisted tunneling  for TE and TM coupling. c) Resonant responsivity of grating-coupled tunnel diode.  Photocurrent versus QCL laser power at fixed resonance $\lambda = 7.3 \mu$m. }
\label{fig:5}
\end{figure*}
The 2nd order term in the current density expansion in Eq.\ref{eq:current1} contains a zero-frequency dc component, and 2nd harmonic terms at $\omega= \pm 2\omega_0$.     The real valued dc tunnel current results in optical rectification, which is the conversion of high-frequency optical fields into a dc response and is given by
\be
J^{(2)}_{dc}   =    \frac{2e }{\hbar} \frac{\pi}{2}  \left(\frac{e\hbar}{m \omega_0}\right)^2  E_0^2 \big(\text{Im}  F^{(2)}(\omega_0)+ \text{Im}F^{(2)}(-\omega_0) \big),
\label{eq:dccoh}
\ee
where we use the imaginary part of the form-factors.   Explicitly , we can evaluate the 2nd order form-factor in the uniform barrier picture.    The resulting real valued dc current  density is given by
\begin{widetext}
\ba
J^{(2)}_{dc}  &  =   & AT^2  \left(\frac{e E_0 t_{ox}}{\hbar  \omega_0}\right)^2  \left( \frac{m_q m_k}{m^2} \right) \frac{\beta}{2} \int dE_k    \left\{   \tau^2(E_k,E_k-\hbar\omega_0)  \ln\left( \frac{1 + \exp(-\beta(E_k-\hbar\omega_0)}{1 + \exp(-\beta E_k )} \right) \right.  \nonumber \\
& +&  \left.\tau^2(E_k,E_k+\hbar\omega_0) \ln\left( \frac{1 + \exp(-\beta(E_k+\hbar\omega_0)}{1 + \exp(-\beta E_k )} \right) \right\}, \label{eq:Jcoherent}
\ea
\end{widetext}
where $A$ is Richardson's constant, $\beta=1/k_b T$, and $\tau(E,E+\hbar \omega_0)$ is the tunneling transmission.  Figure \ref{fig:5}a shows the experimental photocurrent measurement setup  of a large area unipolar grating-coupled tunnel diode measured under coherent quantum cascade laser (QCL)  illumination.  A wavelength tunable QCL with polarized output is incident on a polarization control section followed by a beam-splitter and power monitoring  section with the main beam incident on the sample.  An optional chopper with Zurich lock-in can be used to observe small signal voltages.  The measured responses were obtained without the optional chopper and lock-in and calibrated over wavelength, polarization and optical power. The  photocurrent is measured at zero applied bias and the illumination spot size significantly under fills the grating-coupled diode area. The fabricated grating-coupled tunnel diode is detailed elsewhere \cite{Davids:2015aa,PhysRevApplied.9.054040}.  The spectral photoresponse shown if Fig.\ref{fig:5}b for the TM illumination shows resonantly enhanced photocurrent indicative of photon-assisted tunneling from the confined infrared optical field in the tunnel barrier.  The TE polarization parallel to the metal grating shows no resonant response in this wavelength range. This agrees with the simulated and measured reflection mapping of the n+ metal-oxide silicon grating-coupled tunneling devices\cite{Davids:2015aa,PhysRevApplied.9.054040}.   Furthermore, Fig. \ref{fig:5}c shows the linear response of the device at resonance  as a function of the QCL input power.  This linear scaling with power $P\propto E_0^2$ is predicted by  Eq.\ref{eq:dccoh} and allows for the direct measurement of the 2nd order conductance.   The slope of the TM rphotoresponse versus incident optical power  (Fig. \ref{fig:5}c) relates the differential responsivity $R = dI/dP$ to the second order conductivity,  $2Z_0\sigma^{(2)} = 0.7\mu$A/W .  

 Similarly, the real-valued 2nd harmonic current density is 
\be
J^{(2)}(\pm 2\omega_0)  =    \frac{2e }{\hbar} \frac{\pi}{2}  \left(\frac{e\hbar}{m \omega_0}\right)^2  E_0^2 ~ \text{Im}\left( F^{(2)}(\pm\omega_0) \right)
\ee
where $\omega_0$ is the coherent photon frequency.    The 2nd harmonic photon-assisted tunneling current occurs at $\lambda_{2nd} = 3.65\mu$m which occurs in the region where the heavily doped Si high index substrate is absorptive and semi-transparent\cite{ginn2011infrared}, which makes measurement of 2nd harmonic term difficult in an FTIR spectrometer.    Moreover, it is important to note that the form-factors are in general complex and that the current density expansion from 2nd order and higher  order terms contain imaginary components proportional to the principle value integration.  

\section{Summary}\label{sec4}

In this paper, an expansion of the dynamic photon-assisted tunneling current is presented that arises from a perturbative expansion of the density matrix for a broadband electromagnetic field confined in the barrier of an asymmetric tunneli diode.  The non-linear frequency dependent current density is expanded in terms of the  non-linear  conductances where the power law dependence on the electric field at nth order is due to a corresponding multi-photon (n-1) tunneling process.  A detailed examination of the 2nd order current density shows that a strong dc current response leads to a subsequently large second harmonic response.  The second order response maybe difficult to observe since the radiation is preferentially radiated into the high index lossy dielectric Si substrate at the 2nd harmonic waevelength.  Furthermore, it may be possible to mix two infrared QCL sources in the resonance bandwidth of the grating-coupled tunnel diode  to generate a THz voltage signal on the MOS tunnel device.  Third order harmonics and higher order terms are seen to give small corrections to lower order dynamic current expressions.   While 2nd harmonic generation at metallic surfaces has been predicted and observed\cite{PhysRevB.33.3688,PhysRevLett.46.145,PhysRevLett.63.318}, the dc response of the enhanced grating-coupled diode shows the impact on the resonant ENZ field concentration on the photon-assisted tunnel generation of the non-linear response.  Moreover, by designing the antenna-coupled tunnel diode response to longer wavelengths, the dispersive loss effects of the substrate can be mitigated and the 2nd harmonic signal increased.

\acknowledgements{
Funding for this work was provided by Sandia's Laboratory Directed Research and Development
(LDRD) program. Sandia National Laboratories  is a multi-mission laboratory managed and operated
by National Technology and Engineering Solutions of Sandia, a wholly owned subsidiary of Honeywell International Inc., for the United States Department
of Energy's National Nuclear Security Administration under contract DE-NA0003525. 

This paper describes objective technical results and analysis. Any subjective views or opinions that might be expressed in the paper do not necessarily represent the views of the U.S. Department of Energy or the United States Government.
}

\bibliography{../../bibliography/bibliography}

\appendix
\begin{widetext}
\section{Uniform Barrier: Form-Factors}\label{app:1}
The evaluation of the form-factors in equation \ref{ff1} result from the uniform barrier matrix elements. The wavefunctions in the barrier from the left and right regions are given by 
\ba
\phi_k&  = & \sqrt{1/V}\exp(i\mathbf{k}\cdot \mathbf{r} -k x)\\
\psi_q &=&  \sqrt{1/V}\exp(i\mathbf{q}\cdot \mathbf{r}-q (t_{ox} - x))
\ea
where $\mathbf{k}$ and $\mathbf{q}$ are the transverse momentum.  We convert the summations to integration over momenta
\ba 
\sum_n & \rightarrow V \int \frac{dk}{(2\pi)}  \int \frac{d\mathbf{k}}{(2\pi)^2}  \\
\sum_m & \rightarrow V \int \frac{dq}{(2\pi)}  \int \frac{d\mathbf{q}}{(2\pi)^2} ,
\ea
and the matrix elements, $u_{kq} = t_{kq} (2 \pi)^2 \delta(\mathbf{k} - \mathbf{q})/V$ where $V$ is the quantization volume and comes from the wavefunction normalization.  The energy is given by 
\be
E_n \rightarrow E_k + \frac{\hbar^2}{2m} \mathbf{k}^2,
\ee
with a similar expression for $E_m$ with $\mathbf{k} \rightarrow \mathbf{q}$. We also assume that the distribution function $f_n$ is the Fermi function and we have
\be
f_n \rightarrow f_{k,\mathbf{k}} =   \frac{2}{\exp(\beta(E_k + \frac{\hbar^2}{2m} \mathbf{k}^2)) +1},
\ee
where $\beta = 1/k_b T$  and $T$ is the carrier temperature. 
The resulting form-factors can be computed by integrating over the transverse momentum to give,
\ba
F^{(1)} & = &  \frac{m}{\pi\beta \hbar^2} \int \frac{dq}{(2\pi)} t_{qq} \ln( 1 + \exp(-\beta E_q)) \\
F^{(2)}(\omega_1) &  = &  \frac{m}{\pi\beta \hbar^2}   \int \frac{dk}{(2\pi)} \int \frac{dq}{(2\pi)}    t_{qk} t_{kq}  G_{kq}(\omega_1)   \ln\left( \frac{1 + \exp(-\beta E_q)}{1 + \exp(-\beta E_k )} \right)   \\
F^{(3)}(\omega_1, \omega_2) &  = &  \frac{m}{\pi\beta \hbar^2} \int \frac{dk}{(2\pi)} \int \frac{dq}{(2\pi)}  \int \frac{dK}{(2\pi)}t_{qk} t_{kK}  t_{Kq} G_{kq}(\omega_1)  \left\{ G_{Kq}(\omega_2)  \ln\left( \frac{1 + \exp(-\beta E_q) }{1 + \exp(-\beta E_K ) } \right)  \right.  \nonumber \\
& & \left. - G_{kK}(\omega_2)  \ln\left( \frac{1 + \exp(-\beta E_K) }{1 + \exp(-\beta E_k ) } \right) \right\} .
\ea
The Greens function  is given as 
\be
G_{kq} (\omega)  = \frac{1}{\hbar \omega -E_k + E_q},
\ee
which is just the perpendicular energy denominator.   We define the energy difference, $\Delta_{kq} = E_k -E_q$ and the principal value for the Greens function is 
\ba
G_{kq} (\omega)  & =  & \frac{1}{\hbar \omega -\Delta_{kq} \pm i\eta} \nonumber \\
                            & =  &{ \cal P} \left(  \frac{1}{\hbar \omega -\Delta_{kq}} \right) \mp  i\pi \delta (\hbar \omega -\Delta_{kq}).
\ea
We recall that the matrix elements for transfer interaction are evaluated in the barrier of thickness $t_{ox}$ and  are given by
$t_{qk} = -kt_{ox} \tau_{qk}$, and $t_{kq} = q t_{ox} \tau_{kq}$,
where $\tau$ is the  symmetric barrier transmission and $k(q)$ is the perpendicular momentum.  The barrier transmission is 
\be
\tau_{qk} =  e^{-(k +q)t_{ox}/2} \text{sinhc}\left( (k-q)t_{ox}/2 \right)
\ee
The perpendicular momentum in the tunnel barrier is related to the perpendicular energy where $E_{k,q} <\mathbf{V}$ the barrier height  through the definition 
\be
k = \sqrt{\frac{2m_k}{\hbar^2} \left(\mathbf{V}-E_k \right) },
\ee
and
\be
q = \sqrt{\frac{2m_q}{\hbar^2} \left(\mathbf{V}-E_q \right) },
\ee
where $m_{k,q}$ are the perpendicular effective masses on either the right or left-hand side of the barrier.    The form-factors can be expressed as integrals over the barrier energy, and is given by
\ba
F^{(1)} & = &  \frac{m t_{ox}}{\pi\beta \hbar^2}\left( \frac{m_q}{\hbar^2} \right) \int \frac{dE_q}{(2\pi)} \tau_{qq} \ln( 1 + \exp(-\beta E_q)) \\
F^{(2)}(\omega_1) &  = &  \frac{m t^2_{ox} }{\pi\beta \hbar^2} \left( \frac{m_q m_k}{\hbar^4} \right)  \int \frac{dE_k}{(2\pi)} \int \frac{dE_q}{(2\pi)}    \tau_{qk} \tau_{kq}  G_{kq}(\omega_1)   \ln\left( \frac{1 + \exp(-\beta E_q)}{1 + \exp(-\beta E_k )} \right)   \\
F^{(3)}(\omega_1, \omega_2) &  = &  \frac{m t^3_{ox} }{\pi\beta \hbar^2}  \left( \frac{m_q m_km_K}{\hbar^6} \right)  \int \frac{dE_k}{(2\pi)} \int \frac{dE_q}{(2\pi)}  \int \frac{dE_K}{(2\pi)}\tau_{qk} \tau_{kK}  \tau_{Kq} G_{kq}(\omega_1)  \Big\{  \  \nonumber \\
& &  G_{Kq}(\omega_2)  \ln\left( \frac{1 + \exp(-\beta E_q) }{1 + \exp(-\beta E_K ) } \right) - G_{kK}(\omega_2)  \ln\left( \frac{1 + \exp(-\beta E_K) }{1 + \exp(-\beta E_k ) } \right) \Big\}
\ea
The second order form-factor can be rewritten using the principal value expansion for the Green's function to give break up into real and imaginary parts, we have
\be
 \text{Re}\left(F^{(2)}(\omega) \right)   =  \frac{m t^2_{ox}}{\pi\beta \hbar^2} \left( \frac{m_q m_k}{\hbar^4} \right)  \int \frac{dE_k}{(2\pi)} \int \frac{dE_q}{(2\pi)}    \tau_{qk} \tau_{kq}    \ln\left( \frac{1 + \exp(-\beta E_q)}{1 + \exp(-\beta E_k )} \right)  { \cal P} \left(  \frac{1}{\hbar \omega -\Delta_{kq}} \right)  
 \ee
 \be
\text{Im}\left(F^{(2)}(\omega) \right)   =   \mp  \frac{m t^2_{ox}}{\beta \hbar^2} \left( \frac{m_q m_k}{\hbar^4} \right)  \int \frac{dE_k}{(2\pi)} \int \frac{dE_q}{(2\pi)}    \tau_{qk} \tau_{kq}    \ln\left( \frac{1 + \exp(-\beta E_q)}{1 + \exp(-\beta E_k )} \right)  \delta (\hbar \omega - E_k + E_q).
\ee
We can perform the energy integral and obtain
\be
\text{Im}\left(F^{(2)}(\omega) \right)  =  \mp   \frac{m t^2_{ox}}{2\pi\beta \hbar^2} \left( \frac{m_q m_k}{\hbar^4} \right)  \int \frac{dE_k}{(2\pi)}     \tau^2(E_k,E_k-\hbar\omega)  \ln\left( \frac{1 + \exp(-\beta(E_k-\hbar\omega)}{1 + \exp(-\beta E_k )} \right).
\ee

\end{widetext}

\end{document}